\documentclass[twocolumn,aps,tightenlines,floatfix,showpacs]{revtex4}
\usepackage{graphicx}
\begin{document}

\title{The Nuclear Pairing Gap - How Low Can It Go?}

\author{B. Alex Brown}

\affiliation{National Superconducting Cyclotron Laboratory and Department
of Physics and Astronomy, Michigan state University, East Lansing,
Michigan 48824-1321, USA}

\begin{abstract}
The pairing gap for $^{53}$Ca obtained from
new experimental data on the masses of $^{52-54}$Ca has
the smallest value yet observed. This is explained in the
framework of the nuclear shell model with schematic and realistic
Hamiltonians as being due to shell gaps around the low-$  j  $ orbital $  1p_{1/2}  $.
Minima in the pairing gaps for all nuclei are shown and discussed.
\end{abstract}

\pacs{21.10.Dr, 21.30.-x, 21.60.Cs}

\maketitle

\section{Introduction}

One of the most robust signatures of pairing in nuclei is the odd-even
oscillation in the one-neutron separation energies as a
function of neutron or proton number.
This is illustrated in Fig. 1 which shows the binding energies
and one-neutron
separation energies for the calcium isotopes.
The figure also shows the
results of a shell-model calculation in the
$  (0f_{7/2},0f_{5/2},1p_{3/2},1p_{1/2}) (fp)  $ model space with the GX1A
Hamiltonian (also referred to as GXPF1A in the literature \cite{gx1a})
compared with experiment.
The oscillation in the one-neutron separation energies
can be quantified
in terms of the energy differences
$$
D_{n}(N) =  (-1)^{N+1} [S_{n}(Z,N+1)-S_{n}(Z,N)]
$$
$$
= (-1)^{N} [2 {\rm BE}(Z,N) - {\rm BE}(Z,N-1) - {\rm BE}(Z,N+1)],
$$
where $  S_{n}(N) = {\rm BE}(Z,N) - {\rm BE}(Z,N-1)  $
is the one-neutron separation energy.
$  N  $ is the number of neutrons and $  Z  $ is the
number of protons. This quantity turns out to be always positive and reflects
the fact that the even nuclei are always more bound on the average
that the neighboring odd nuclei. I will distinguish the results
for even and odd $  N  $ values denoted by, $  D_{ne}  $ and $  D_{no}  $, respectively.
In the literature one
commonly finds the related quantity known as the odd-even mass
parameter or pairing gap
$  \Delta _{n}(N) = \frac{D_{n}(N)}{2}  $ (see Fig. 2.5 in \cite{bm}).
I use $  D  $ rather than $\Delta$ because its values are more directly
connected to simple underlying quantities associated with pairing and shell gaps.
Equivalent equations as a function of proton number
are obtained by fixing $  N  $ and varying $  Z  $.
\begin{figure}
\includegraphics[scale=0.5]{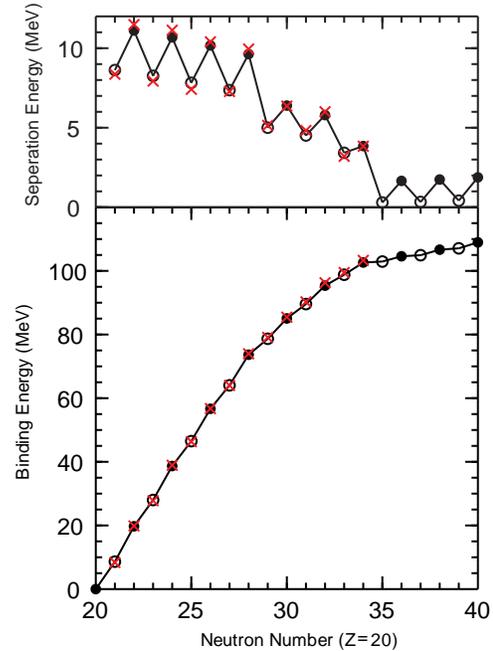}
\caption{The bottom panel shows the ground-state
energies for the calcium isotopes
obtained from the $  pf  $ shell-model calculation with the GX1A
Hamiltonian relative to $^{40}$Ca with
filled circles even $  N  $ and open circles for odd $  N  $,
all connected by a line. The
crosses are the experimental data. The top panel shows
the one-neutron separation energies for GX1A and experiment.}
\end{figure}
\begin{figure}
\includegraphics[scale=0.5]{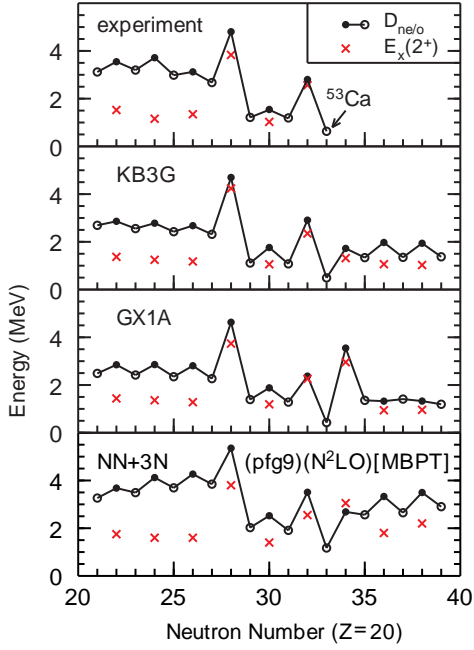}
\caption{
$  D  $ and E$_{{\rm x}}$(2$^{ + }$) for the calcium isotopes as a function
of neutron number. $  D  $ values are shown by the filled circles
for even $  N  $ and open circles for odd $  N  $, all connected
by a line. The E$_{{\rm x}}$(2$^{ + }$) values are shown by the crosses.}
\end{figure}

Fig. 2 shows values of $  D_{n}  $ for the calcium isotopes ($  N>20  $) obtained from
experiment and from two commonly used effective Hamiltonians in the $  fp  $
model space, KB3G \cite{kb3g} and GX1A \cite{gx1a} and with a new ab-initio
Hamiltonian that includes three-nucleon interactions \cite{mbpt}.
I also show the excitation energies for the lowest 2$^{ + }$ states of the
even nuclei. The experimental data is from the 2012 mass table \cite{wang}
together with recent data for $^{51-54}$Ca \cite{nature}.
The new data provide a point for
$^{53}$Ca in the upper panel of Fig. 2 that turns out to the
the lowest value for $  D_{n}  $ observed in all nuclei with even $  Z  $.
The results are in good agreement with the KB3G and GX1A shell model
predictions. In this letter I use the shell model with
schematic and realistic Hamiltonians
to understand the trends observed for $  D  $, and in particular the
low value for $^{53}$Ca. This will be used to qualitatively
understand the trends for minima in the $  D  $ values for all nuclei.

To obtain insight into the reasons for the patterns observed
in Fig. 2, I start with the simple ``surface-delta-function" (SDI)
model for the interaction \cite{plas66}. The SDI differs from the
delta interaction by the replacement of the radial integrals
by a constant. The results for $  D  $ obtained with SDI when the single-particle
energies are degenerate are shown in the
bottom panel of Fig. 3. The SDI $  J=0  $, $  T=1  $ two-body matrix elements for
orbitals $  a  $ and $  b  $ with spins $  j_{a}  $ and $  j_{b}  $ are
$  <a a\mid  {\rm SDI} \mid  b b>\, = C \sqrt{(2j_{a}+1)(2j_{b}+1)}  $,
where $  C  $ is a constant.
The interaction strength $  C  $
is chosen to give a value for $  D  $ that will turn out to
be similar to that obtained with the effective $  pf  $ shell
Hamiltonians for calcium. The excitation energies of the 2$^{ + }$ states
are also constant with the SDI and degenerate single-particle
energies. The results for $  D  $ would be the same if the
$  pf  $ orbitals were replaced by a single orbital with $  j=19/2  $,
but the constant 2$^{ + }$ energy would be higher (3 MeV). The $  D  $ value
is determined by the number of $  m  $ states that
participate in the pairing.
The
interaction energies
obtained with SDI are  $  E(n) = \frac{n V_{o}}{2}  $ for even $  n  $,
and $  E(n) = \frac{(n-1)V_{o}}{2}  $ for odd $  n  $,
and thus $  D = - V_{o}  $. $  V_{o}  $ is the paired interaction
strength for two particles with $  J=0  $.
All of the odd $  N  $ nuclei have four
degenerate states with $  J^{\pi }  $ = 1/2$^{-}$, 3/2$^{-}$, 5/2$^{-}$ and 7/2$^{-}$.
\begin{figure}
\includegraphics[scale=0.5]{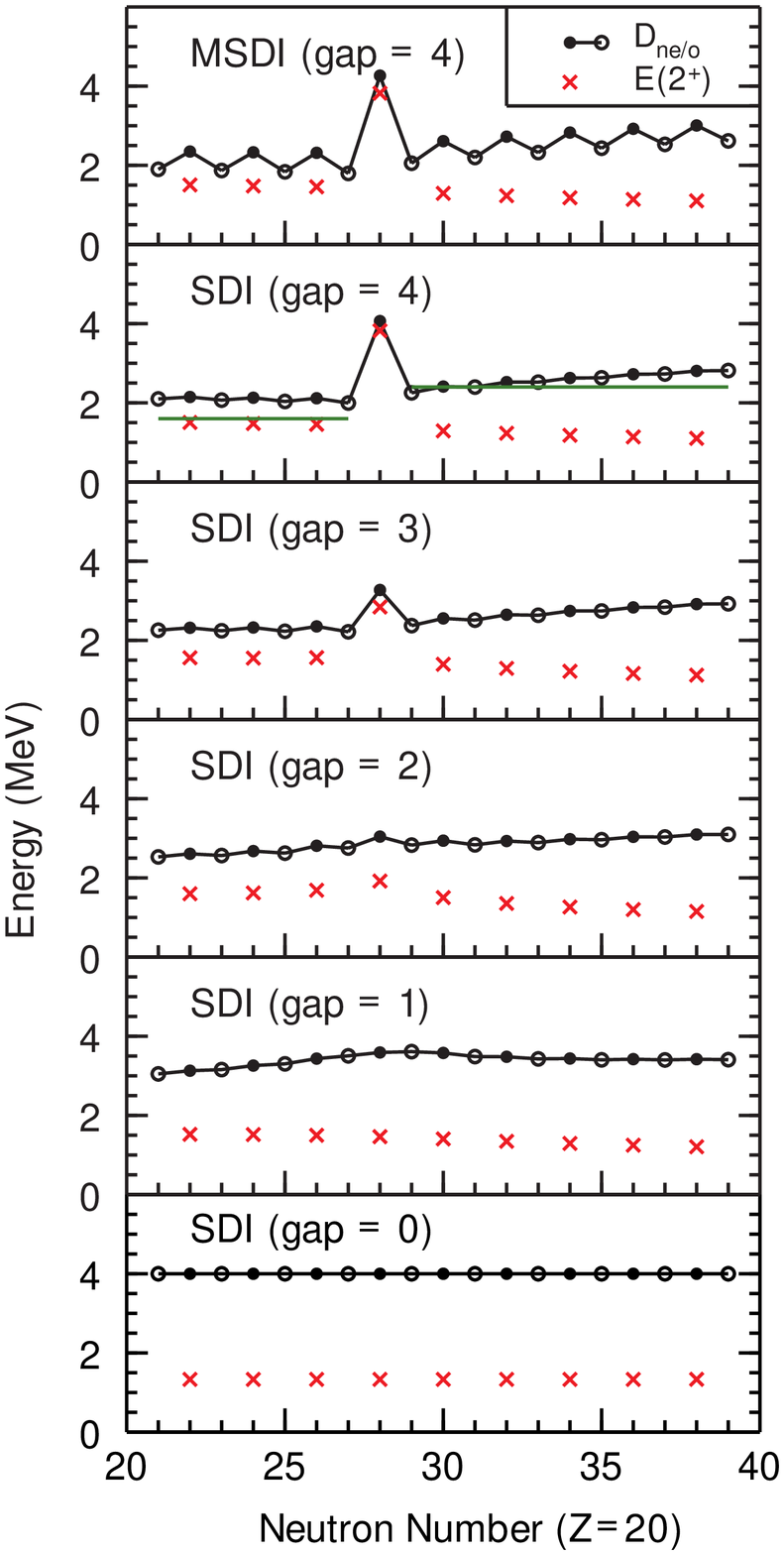}
\caption{Results obtained with the SDI interaction
as a function of the shell gap between the $  0f_{7/2}  $
and $  (0f_{5/2},1p_{3/2},1p_{1/2})  $ set of orbitals.
The top panel also show the results for $  D  $ (thin lines)
obtained when the shell gap is infinite.}
\end{figure}
\begin{figure}
\includegraphics[scale=0.5]{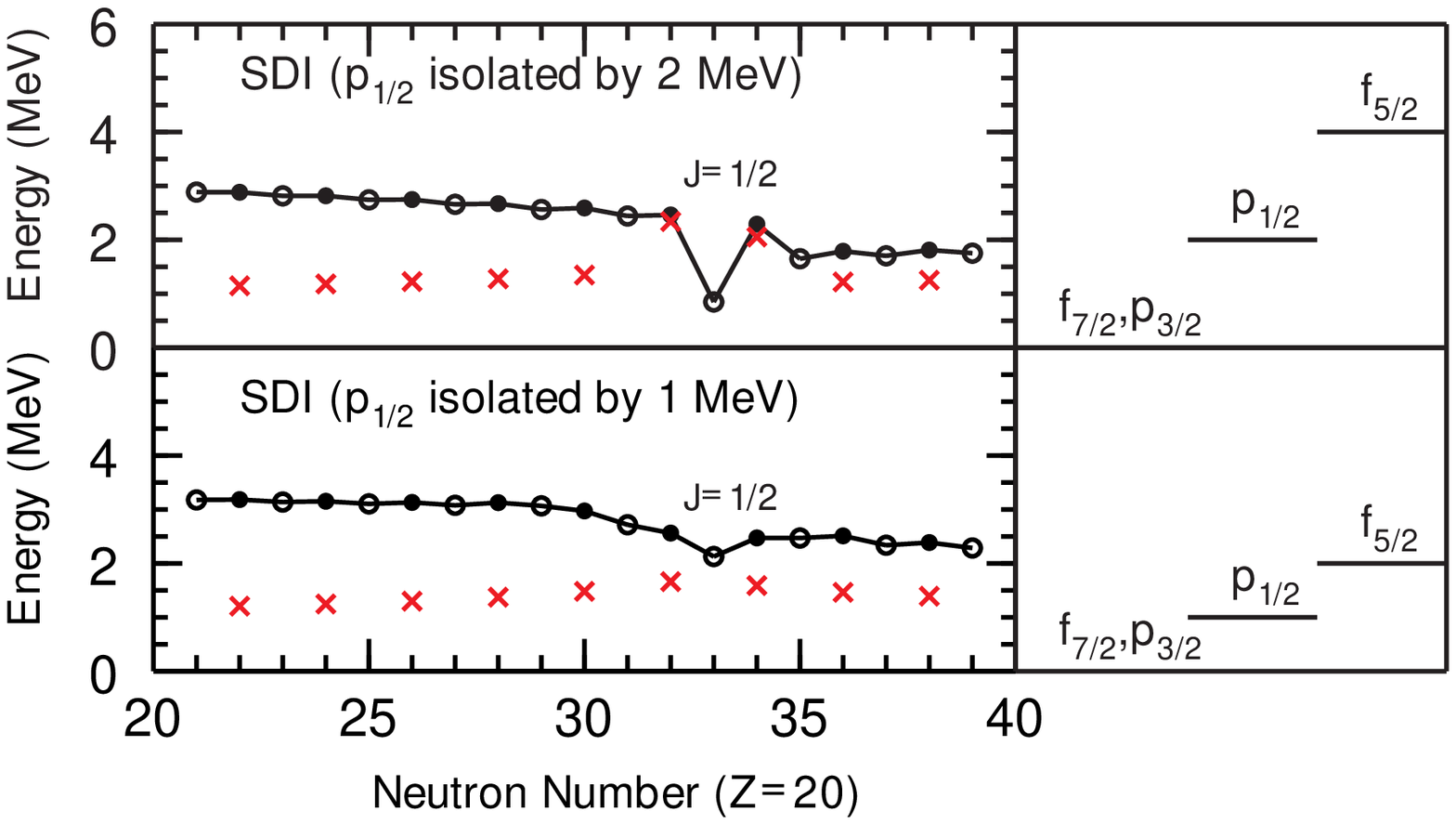}
\caption{Results obtained with the SDI interaction
as a function of the shell gap between the $  0f_{7/2}  $
and $  (0f_{5/2},1p_{3/2},1p_{1/2})  $ set of orbitals.
The top panel also show the results for $  D  $ (thin lines)
obtained when the shell gap is infinite.}
\end{figure}

Fig. 3 shows the numerical results obtained as the shell
gap between the $  0f_{7/2}  $ orbital and a degenerate
group of $  0f_{5/2},1p_{3/2},1p_{1/2}  $ orbitals is increased from
zero to four MeV. The value of $  D_{e}  $ at $  N=28  $ begins to rise
when the value of shell gap becomes greater than $  D_{o}  $.
In the infinite gap limit the total original value of 4 MeV
is divided between the lower group (1.6 MeV for the $  0f_{7/2}  $)
and the upper group (2.4 MeV for $  0f_{5/2},1p_{3/2},1p_{1/2}  $).
When the shell gap is well formed the value of $  D  $ is nearly equal
to that of the 2$^{ + }$ excitation energy. The experiment and
calculations in Fig. 2 show clear signatures of shell gaps
at $  N=28  $ and $  N=32  $. The GX1A calculation also shows a
shell gap at $  N=34  $ in contrast to the KB3G and NN$+$3N models
that do not show a gap at $  N=34  $. Mass measurements
for the more neutron-rich calcium isotopes are required
for the experimental value at $  N=34  $. Indirect
evidence from prodiction cross section data indicates that
the the shell gap at $  N=34  $ is about 0.5 MeV smaller
than that given by GX1A (and GX1B) \cite{frag}.

In order to obtain a dip in the $  D_{o}  $ value one needs to
make shell gaps below and above a low-$  j  $ orbital. This
is shown in Fig. 4 where there are three groups of
orbitals with $  1p_{1/2}  $ in the middle split by one and two MeV
from the other orbitals.
In the limit of a completely isolated $  1p_{1/2}  $ orbital
$  D_{o} = - C \sqrt{(2j_{a}+1)(2j_{b}+1)} = - 2 C = 0.4  $ MeV.

The main defect of the SDI model is the lack of oscillations
in the $  D  $ values that are observed in experiment.
In the 1960's
this was recognized as a basic failure of the delta and SDI Hamiltonians.
It was empirically fixed by adding a constant to the interaction to make
the so-called modified-delta interaction \cite{glaud67}
of the form $  V(\mid \vec{r}_{1}-\vec{r}_{2}\mid ) = A \delta (\mid \vec{r}_{1}-\vec{r}_{2}\mid ) 
+ B  $.
With the SDI form of the radial integral this becomes
the so-called modified-surface-delta interaction (MSDI).
The modern interpretation of this constant is that it comes from
core-polarization corrections and three-body interactions.
It is essential to obtain a good saturation property for
the binding energies.
The $  D  $ values obtained with $  B=0.2  $ MeV and
with a shell gap of four MeV are shown
at the top of Fig. 3.
This constant simply adds a term $  n(n-1)B/2  $
to the all energies and $  D_{o} = -V_{o} - B  $ for odd $  n  $ and
$  D_{e} = -V_{o} + B  $ for even $  n  $.
Half the sum of neighboring even and odd $  D  $
gives the pairing contribution:
$  D_{a} = \frac{1}{2}[D_{e}(N)+D_{o}(N-1)] = -V_{o}  $,
and half of the difference
gives quadratic dependence: $  D_{b} = \frac{1}{2}[D_{e}(N)-D_{o}(N-1)] = B  $.
Fig. 2-5 in Bohr and Mottelson \cite{bm} is based upon the
$  D_{a}  $ combination ($  \Delta  = D_{a}/2  $).

The Coulomb interaction between protons behaves
like the addition of a long-range monopole term
In the liquid drop
model $  B = (6e^{2}/5R)  $. There is also a small
anti-pairing addition to $  D_{a}  $ of (about $+$0.10 MeV for $  Z=20  $).
This is useful for understanding the trends in $  D_{p}(Z)  $
compared to those for $  D_{n}(N)  $.

Thus, the MSDI model with shell gaps gives a semi-quantitative
understanding of all trends observed in $  D_{e/o}  $.
The low $  D_{o}  $ value for $^{53}$Ca observed in experiment
and theory in Fig. 2 is due to occupation of
the $  1p_{1/2}  $ orbital at $  N=33  $. The experimental value for
$^{53}$Ca is $  D_{o}=0.63(10)  $ MeV compared to the calculated
values (in MeV) of 1.170 (MBPT), 0.425 (GX1A) and
0.489 (KB3G).
The key quantity for
the effective Hamiltonians is the
$  <(1p_{1/2})^{2}\mid  V \mid  (1p_{1/2})^{2}>  $
effective two-body matrix element.
It is 0.151 MeV for KB3G and 0.053 MeV for GX1A
compared to -0.20 MeV with the MSDI model. If the $  1p_{1/2}  $
orbital was completely isolated its $  D_{o}  $ value would
be negative for GX1A and KB3G.
The small positive $  D_{o}  $ values obtained
with GX1A and KB3G is due to mixing with the other
orbitals.
\begin{figure}
\includegraphics[scale=0.6]{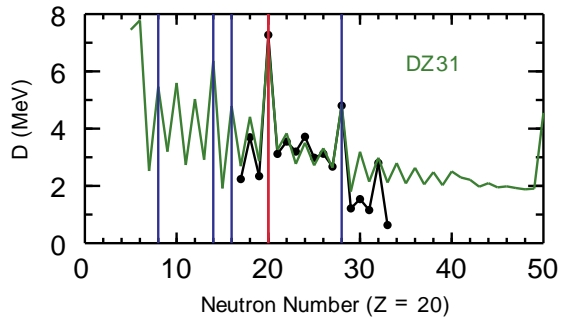}
\caption{$  D  $ value for the calcium isotopes from experiment
(points connected a line) compared to the DZ31 mass model \cite{dz31} (green line).}
\end{figure}
\begin{figure}
\includegraphics[scale=0.6]{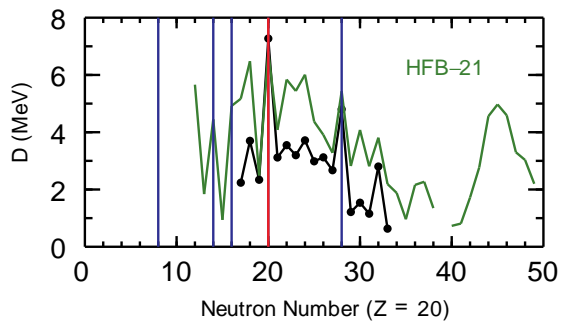}
\caption{$  D  $ value for the calcium isotopes from experiment
(points connected by a line) compared to those obtained with the
HFB-21 Skyrme interaction \cite{hfb21} (green line).}
\end{figure}

To go beyond the $  pf  $ shell we need to consider global mass
models such as the 31 parameter model of Duflo and Zuker (DZ31) \cite{dz31}, or
the results from more microscopic calculations such as the
energy-density functional method with the HFB-21 Skyrme interaction \cite{hfb21}.
The DZ31 results shown in Fig. 5 are in rather good agreement with
experiment, but it misses the dip at $  N=33  $ Some parameters
of the DZ31 model are determined from this region, but grouping of
orbitals considered for the pairing is not correct above $  N=28  $.
The HPF-21 predictions shown in Fig. 6
are rather poor. This could be due to
incorrect single-particle energies or an incorrect pairing interaction.

\begin{figure}
\includegraphics[scale=0.5]{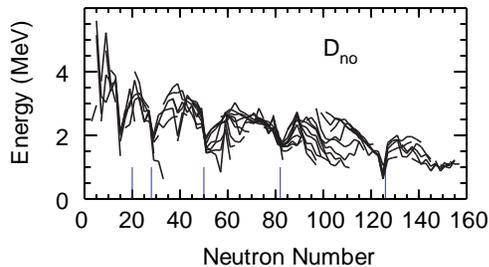}
\caption{Experimental values of $  D_{no}  $ for all nuclei
with even $  Z  $. Values are plotted as a function of the number
of neutrons $  (N>Z)  $ and connected by lines for a given $  Z  $ value.
The vertical lines show the location of the
magic numbers 20, 28, 50, 82 and 126.}
\end{figure}
\begin{figure}
\includegraphics[scale=0.5]{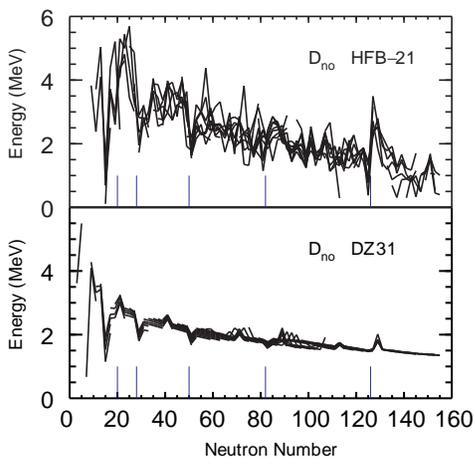}
\caption{Theoretical values of $  D_{no}  $ for all nuclei
with the same range of $  N  $ and $  Z  $ given by the experimental
data in Fig. 7.}
\end{figure}
\begin{figure}
\includegraphics[scale=0.5]{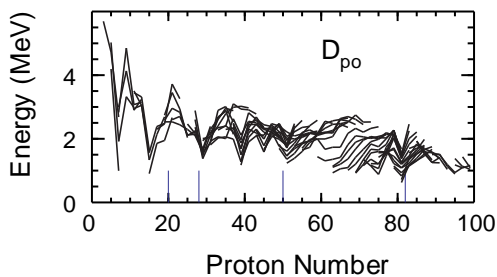}
\caption{Experimental values of $  D_{po}  $ for all nuclei
with even $  N  $. Values are plotted as a function of the number
of protons $  (N>Z)  $ and connected by lines for a given $  N  $ value.
The vertical lines show the location of the
magic numbers 20, 2 and 50, 82.}
\end{figure}

The experimental $  D_{no}  $ values for all nuclei obtained
from the 2012 mass table \cite{wang} and the
new calcium experiment \cite{nature} are shown in Fig. 7.
The value for the $^{53}$Ca at $  N=33  $ is the lowest value
obtained for all nuclei and ties with that of $^{207}$Pb that
has a value of 0.630. But if one scales residual interaction
strength roughly as $  (A)^{-1/2}  $, the scaled minimum
is 50\% lower in $^{53}$Ca as compared to $^{207}$Pb.
Another relatively low value occurs for
$  N=57  $ at $^{97}$Zr and is
associated with a relatively isolated $  2s_{1/2}  $
orbital at that point.

The minima at $  N=33  $ is only observed
for calcium ($  Z=20  $) indicating that required shell
gaps quickly disappear. For increasing $  Z  $ the $  1p_{1/2}  $
and $  0f_{5/2}  $ orbitals cross over, and at $  N=39  $
there is a small minimum coming from small gaps
around the $  1p_{1/2}  $ orbital in $^{68}$Ni ($  Z=28  $)
and $^{70}$Zn ($  Z=30  $).
Robust minima are observed for $  N=15  $ due to the
$  1s_{1/2}  $ orbital and $  N=127  $ due to the $  2p_{1/2}  $
orbital. All of these nuclei have a $  J=1/2  $ spin.
The $  0d_{5/2}  $ - $  1s_{1/2}  $ gap disappears for
carbon ($  Z=6  $) \cite{c16}, and the $  N=15  $ dip should be gone
for $^{21}$C. Mass measurements of $^{21,22}$C are required.

The dip at $  N=57  $ is isolated at $^{97}$Zr.
As $  Z  $ increases the $  2s_{1/2}  $ orbital crosses
over the $  0g_{7/2}  $ orbital and creates a small
dip at $  N=65  $ in $^{115}$Sn.  As $  Z  $ decreases
the $  2s_{1/2}  $ orbital should cross over $  1d_{5/2}  $
orbital creating a minimum at $  N=51  $ just
below $^{78}$Ni. Mass measurements of $^{78-80}$Ni are required.

These experimental trends have been noted previously and compared
some HFB models in \cite{bertsch2009}.
The $  D_{no}  $ values obtained from the DZ31 and HFB-21 mass
models are shown in Fig. 8 where they are plotted for the same
range of $  Z  $ and $  N  $ values known experimentally in Fig. 7.
The analytic behaviour of DZ31 is much smoother than experiment.
The HFB-21 results are more chaotic than experiment, but show
some similarities in the location of the dips to experiment.
Niether of these models predicted
a dip at $  N=33  $.
The complete set of comparisons experiment for these models is shown on
my websige \cite{web}.

The experimental $  D_{po}  $ values for all nuclei obtained
from the 2012 mass table \cite{wang} are shown in Fig. 9.
There are robust minima at $  Z=7  $ (nitrogen) due to the $  0p_{1/2}  $
orbital, at $  Z=15  $ (phosphorus) due to the $  1s_{1/2}  $ orbital,
and at $  Z=81  $ (thallium) due to the $  2s_{1/2}  $ orbital.
The robust minimum at $  Z=29  $ (copper) is due to the isolated $  1p_{3/2}  $
orbital. The minimum at $  Z=39  $ (yttrium) starts at $  N=38  $
and is due to the $  2p_{1/2}  $.
orbital. $^{87-97}$Y all have 1/2$^{-}$ ground state spins.
Generally the shell gaps are smoothed out by the energy splitting
of the Nilsson orbitals in deformed nuclei. But the $  Z=39  $ dip remains
in $^{99-101}$Y which are presumably deformed and have
uncertain ground-state spins.

In summary, I have shown the oscillations
in the neutron separation energy are described by the modified
surface delta interaction. In particular the $  D_{no}  $ value
obtained for $^{53}$Ca is the smallest value yet measured
and it due to an isolated $  0p_{1/2}  $ orbital. Other mass regions
where the effects of isolated low-$  j  $ orbitals were shown.

{\bf Acknowledgements:}
I acknowledge support from NSF grant PHY-1068217.

\end{document}